\documentclass[10pt]{article}
\usepackage{amssymb}
\usepackage{graphics}
\usepackage{amsmath}
\usepackage{amsfonts}
\usepackage{graphicx}
\usepackage{multirow}
\usepackage{subfigure}
\usepackage{multirow}
\usepackage{color}

\begin{document}

\title{Self-consistent field predictions for quenched spherical biocompatible triblock copolymer micelles}
\author{J{\' e}r{\^ o}me G.J.L. Lebouille$^{1,2}$, Remco Tuinier$^{2,3}$, Leo F.W. Vleugels$^2$, \\ Martien A. Cohen Stuart$^4$, Frans A.M. Leermakers$^4$\\ \\
$^{1}$ DSM Biomedical, P.O. Box 18, 6160 MD Geleen, The Netherlands \\
$^{2}$ DSM ChemTech, Advanced Chemical Engineering Solutions (ACES), \\P.O. Box 18, 6160 MD Geleen, The Netherlands.\\
$^{3}$ Van 't Hoff Laboratory for Physical and Colloid Chemistry, \\Debye Institute, Utrecht University, Padualaan 8, 3584 CH Utrecht,\\ The Netherlands.\\
$^{4}$ Laboratory of Physical Chemistry and Colloid Science,\\
Wageningen University, Dreijenplein 6, 6307 HB Wageningen, \\
The Netherlands.
}
\maketitle

\begin{abstract}
We have used the Scheutjens-Fleer self-consistent field (SF-SCF) method to predict the self-assembly of triblock copolymers with a solvophilic middle block and sufficiently long solvophobic outer blocks. We model copolymers consisting of polyethylene oxide (PEO) as solvophilic block and poly(lactic-co-glycolic) acid (PLGA) or poly($\epsilon$-caprolactone) (PCL) as solvophobic block. These copolymers form structurally quenched spherical micelles provided the solvophilic block is long enough. 
Predictions are calibrated on experimental data for micelles composed of PCL-PEO-PCL and PLGA-PEO-PLGA triblock copolymers prepared via the nanoprecipitation method.
We establish effective interaction parameters that enable us to predict various micelle properties such as the hydrodynamic size, the aggregation number and the loading capacity
of the micelles for hydrophobic species that are consistent with experimental finding. 
\end{abstract}

\textbf{keywords:} Micelle, Scheutjens-Fleer Self Consistent Field theory (SF-SCF), block copolymers, encapsulation.

\clearpage

\section{Introduction}
\label{sec:1}

Surfactants, amphiphiles and copolymers in selective solvents are widely used to create micelles. For several applications the size, size distribution, loading capacity and stability (upon dilution) are requirements of great importance. It is feasible to invoke the statistical thermodynamical machinery on invariably coarse grained models to find detailed information on self-assembly phenomena. However, the insights from this have a qualitative rather than a quantitative character because of a lack of suitable parameter sets for such systems. This is why in practice the selection of appropriate copolymers capable of forming micelles with pre-set requirements is still based on trial and error or on experience rather than on theory. Confronting theory with experiments is the only way to improve this situation.

In this paper our focus is on triblock copolymers in a selective solvent (water). Several triblock copolymers have been studied both from a theoretical- \cite{Hurter1993,Lauw2006,Alexandridis1994,Monzen1999,Wang2005,Zhulina2012} as well as from a practical \cite{LebouilleA2011,Ryu2001,Chen2005,Ge2002} perspective. Many of these polymers, such as the poloxamers, were selected to have a finite (reasonably high) solubility in water, implying that if the systems are diluted below their critical micelle concentration (CMC), the micelles break up into freely dissolved unimers. When these micelles are used to encapsulate compounds then, upon dilution below the CMC, their cargo is released. We study biocompatible triblock copolymers that have a very limited water solubility. The micelles may still be used as drug carriers as one can employ alternative release strategies. More specifically, we use copolymers made from PLGA \cite{Anderson1997} or PCL \cite{Sinha2004} as the hydrophobic fragment and PEO as the hydrophilic species. The PLGA and PCL blocks are biodegradable by hydrolysis of the esters and or are subject to enzymatic degradation \cite{Peters2011,Sodergard2002,Nie2003,Hu2004} and the PEO block is mainly excretable via the renal pathway \cite{PEOexcr,Knop2010}.

The block lengths of the PLGA-PEO-PLGA and PCL-PEO-PCL polymers are chosen such that they exhibit ultra-low water solubilities. Experimentally one can reproducibly fabricate micellar objects by the precipitation method and the corresponding micelles may be referred to as frozen or 'dead' because after their formation they do not dissolve by dilution. Our interest is in the modeling of these structurally quenched systems by using an equilibrium self-consistent field (SCF) theory. This is not a trivial exercise because the micellar system clearly violates the important prerequisite of equilibration. We may justify our approach {\it postiori}, because for a particular set of interaction parameters it is found that there exists a good correlation between the predicted micelle structure and experimental observations. 

Association colloids composed of molecules in strongly selective solvents have a densely packed core and a solvated corona. Based on this, particularly in the surfactant literature, the surfactant packing parameter $P=v/(a_{0}\times{l_{c}})$ is used to assess the capability of some amphiphile to form a certain association colloid. Here $v$ is the volume of the hydrophobic block(s) (tails), $l_{c}$ is the length of the tail(s) and $a_{0}$ is the surface area occupied by the polar fragments (head) at the CMC. For $P<1/3$ spherical micelles are preferred, whereas for $1/3<P<1/2$ cylindrical micelles form. In the range $1/2<P<1$  vesicles, for $P\approx 1$ planar bilayers and for $P>1$ reversed spherical micelles are expected \cite{Israelachvilli1976,Israelachvilli1985}. For surfactants that have relatively short tails with little conformational degrees of freedom, the main problem in using the packing parameter concept is to estimate $a_0$. This quantity can be derived from experiments. For polymeric self-assembly there are more hurdles to take. In addition to the issue to know the area per 'head' group, it is important to account for the conformational degrees of freedom of the copolymers. 

In the field of polymer self-assembly it is known that the thermodynamic stability of micelles depend on the size of the core compared to that of the corona \cite{Borisov2002, Zhulina2002}. More specifically, a particular geometry is stable when the corona is large compared to the core. Considering, for example, micelles of which the corona is relatively small, the system is expected to reduce the curvature. In the cylindrical geometry the ratio between surface area and volume is less and the corona chains are forced to stretch outward in the radial direction so that the size of the corona increases compared to the core. With a similar argument one can envision the choice of a system for the lamellar phase. In order to use this insight as a predictive tool, one must get involved in the field of polymer brushes and in particular the physics of curved polymer brushes. The analytical methods to estimate the core and corona sizes is mostly limited to scaling relations which must be somehow calibrated.

By using the SCF theory and evoking a molecularly realistic model of the polymers, we can resolve these problems to a large extend. This gives us the capability to predict structural properties of the micelles for given composition of the copolymers. 

In the following we will first give a brief introduction on the SF-SCF theory for micellisation. Subsequently our results are discussed and compared to experimental data presented in more detail in the companion publication \cite{LebouilleA2011}. In our conclusions we elaborate on the use of an equilibrium theory to describe micelles formed by the precipitation method.

\section{SF-SCF  Theory}
\label{sec:4}
\subsection{Thermodynamic considerations}
\label{sec:4.1}

Micellar solutions are macroscopically homogeneous. First and second law of thermodynamics for homogeneous solutions with $i = 1, 2, \cdots, c$ different molecular components, the total numbers of molecules of the $i^{\rm th}$ component $n_i$, consisting of $c$ components with a chemical potential of all components $\mu_i$, give for the change of the internal energy d$U$  for a homogeneous phase:
\begin{equation}
{\rm d}U=T{\rm d}S-p{\rm d}V+\sum_i^c \mu_i {\rm d}n_i
\label{dU}
\end{equation}
where the sum is over all molecular components, $S$ is the entropy and $V$ the system volume. For systems at a given temperature $T$ and pressure $p$ it is often better to turn to the Gibbs energy $G \equiv U - TS + pV$ and the change in the Gibbs energy reads:
\begin{equation}
{\rm d}G=-S{\rm d}T+V{\rm d}p+\sum_i^c \mu_i {\rm d}n_i
\label{dG}
\end{equation}
Classical thermodynamics cannot account for micellisation as specific correlations between molecules are fully acceptable from a thermodynamic perspective and the equations in the presence or in the absence of some finite size aggregates are completely the same. Equations only start to be different as soon as macroscopic phase changes occur. 

Motivated by the knowledge that on some microscopic level the system is inhomogeneous, it may be of interest to consider the small system approach advocated by Hill \cite{Hill1991}. Hill suggested that when there is a hidden parameter, here the number of micelles $\mathcal{N}$, there is an intensive quantity $\varepsilon$, which Hill referred to as the sub-division potential, coupled to the number of micelles and the change of the Gibbs energy reads:
\begin{equation}
{\rm d}G=-S{\rm d}T+V{\rm d}p+\sum_i^c \mu_i {\rm d}n_i + \varepsilon {\rm d} \mathcal{N}
\label{dG_}
\end{equation}
From Eq. \ref{dG_} it follows that the sub-division potential is the work (Gibbs energy) needed to increase the number of micelles for given number of molecules, pressure and temperature. In equilibrium the Gibbs energy should be minimized. This must also apply to the dependence of the Gibbs energy on the number micelles:
\begin{equation}
\left ( \frac{\partial G }{\partial \mathcal{N}}\right )_{T,p,\{ n_i\}} = \varepsilon = 0
\label{Gmin}
\end{equation}
and the second derivative of the Gibbs energy with number of micelles should be positive. In words, Eq. \ref{Gmin} expresses that in equilibrium there is no Gibbs energy associated to the formation of micelles and under these conditions Eq. \ref{dG_} returns to Eq. \ref{dG} obviously. Hence the small system approach is consistent with (macroscopic) thermodynamics. The interesting point of the small system approach is that Scheutjens-Fleer Self Consistent Field (SF-SCF) theory considers the system on the small system level and the small system thermodynamics approach becomes meaningful. In these calculations we focus on one micelle in the center of the spherical coordinate system of which we can change the aggregation number (by considering the number of copolymers in the small system). We may use Eq. \ref{Gmin} to select the relevant number of polymers per micelle.
 
Returning to Eq. \ref{dG_} we notice that $G = \sum_i \mu_i n_i + \varepsilon \mathcal{N}$. The Gibbs energy per micelle is thus: $G/\mathcal{N}=\sum_i \mu_i n_i/\mathcal{N} + \varepsilon$ from which it follows that $\varepsilon$ is interpreted as the excess Gibbs energy per micelle. This quantity is also referred to as the grand potential $\Omega$ per micelle. Hence, equilibrium in the SF-SCF protocol for micellisation is defined by the grand potential of the micelle being zero. Below we will return to this issue obviously. 

\subsection{SF-SCF machinery}
\label{sec:4.2}
We use the classical SF-SCF model for self-assembly, which in the context of surfactant micellisation has been presented in the literature several times \cite{Hurter1993,Scheutjens1979,Scheutjens1980,BookFleer}. Here we will only outline the most important features so that the results of the modeling can be discussed properly. We will pay attention to (i) the discretization scheme, (ii) the molecular model, (iii) the optimization of the free energy, (iv) the propagator scheme and (v) the grand potential in the following subsections.

\subsubsection{The discretization}
\label{sec:4.2.1}
The SF-SCF model is lattice based. This means we have to define exactly how the lattice sites are organized. Here and below we focus on the spherical lattice.  We consider lattice sites with linear length $b$ and volume $b^3$. The lattice sites are arranged in lattice layers with spherical topology. Starting with a central point at $r = 0$, we have layers of lattice sites at coordinate $r = 1, 2, \cdots, M$, which are a distance $rb$ away from the center. The number of lattice sites at coordinate $r$ is given by $L(r) = \frac{4}{3} \pi \left (  r^3-(r-1)^3\right)\approx 4 \pi r^2$, where the approximation on the rhs of this equation (which is accurate only for large values of $r$) shows that the number of sites is related to the area of the shell at distance $rb$ from the center. In this coordinate system we need to compute so-called site averages defined by a three-layer average of some spatially varying quantity $\Phi(r)$, for which we use the angular bracket notation defined by:
\begin{equation}
\left \langle \Phi(r) \right \rangle \equiv \lambda_{r,r-1}\Phi(r-1) + \lambda_{r,r}\Phi(r) +\lambda_{r,r+1}\Phi(r+1) 
\label{site}
\end{equation}
In this equation the \textit{a priori} step probabilities account for the geometry 
\begin{eqnarray}
\lambda_{r,r-1}=\lambda \frac{4\pi (r-1)^2}{L(r)}\\
\lambda_{r,r+1}=\lambda \frac{4 \pi r^2}{L(r)}\\
\lambda_{r,r}=1-\lambda_{r,r-1}-\lambda_{r,r+1}
\label{lamdas}
\end{eqnarray}
For a cubic lattice, the limiting values, that is for large values of $r$,  of the step probabilities are $\lambda_{r,r-1}=\lambda_{r,r+1}=\frac{1}{6}$ and $\lambda_{r,r}=\frac{4}{6}$. For small values of $r$ there is curvature information in the transition probabilities. 

The SF-SCF theory makes use of the mean-field approximation. In practice this means that we are going to average various properties over all sites $L(r)$ at a particular coordinate $r$. 

\subsubsection*{The molecular model}
\label{sec:4.2.2}
Scheutjens and Fleer promoted the idea that the polymeric species should be expressed in segments that fit on the lattice. In other words, a coarse-grained description of the polymer chains is implemented. In this approach the polymers are considered as a string of segments with linear length $b$. Let us for convenience number the different molecules with the index $i$ and focus on linear chains of which the segments have ranking numbers $s = 1, 2, \cdots, N$. where $N$ is the total number of segments in the chain. The chain topology is an input for the calculations. This means that we have to specify exactly what the segment type is of each segment. Segment types are generically referred with the index $X$. For example, we may consider the symmetric triblock copolymers $A_{N_A}B_{N_B}A_{N_A}$,  which has segments of type $X = A$ for the ranking numbers $s = 1, 2, \cdots, N_A$  and $s = N_A + N_B +1, \cdots , 2 N_A + N_B$, and $X = B$ for the remaining ones $s =  N_A  +1,\cdots, N_A + N_B$. Besides polymeric species there may also be monomeric compounds in the system. These are treated similar to the chains, yet they have just one segment $s = 1$. Below we will assume that the solvent has a segment type $S$ and is monomeric. 

For convenience we introduce the so-called chain architecture operators
\begin{equation}
\delta _{s,i}^A  = \left| {\begin{array}{*{20}c}
   1 & {\rm when\ segment}\ s\ {\rm of\ molecule}\ i\ {\rm is\ of\ type}\ A  \\
   0 & {\rm otherwise}  \\
\end{array}} \right.
\end{equation}
These values of these operators are fully defined by the input data. 

The target of the SF-SCF equations is to find the equilibrium distribution of all segments and segment types in the coordinate system. The dimensionless concentration of segments of type $X$ at coordinate $r$ is given by the volume fraction $\varphi_X(r)$, which is given by the ratio between the number of segments of type $X$ at this coordinate and the number of sites available: 
\begin{equation}
\varphi_X(r) = \frac{N_X(r)}{L(r)}
\label{phi_X}
\end{equation}
The SF-SCF theory is based on a mean-field free energy expression. This expression features besides the segment volume fractions also segment potentials $u_X(r)$. Physically, the segment potentials should be interpreted as the work needed to bring segment $X$ from the bulk to the coordinate $r$. From this definition it follows that in the bulk the segment potentials are zero. 

\subsubsection{The free energy and the optimization}
\label{sec:4.2.3}
In the calculation, the volume occupied by the $M$ lattice layers and the number of molecules are specified. Hence, the Gibbs energy is the primary thermodynamic potential in the system. Schematically the Gibbs energy $G$ can be presented as
\begin{eqnarray}
G(\{\varphi \},\{u\},\alpha)= \nonumber \\
-\ln Q(\{u\})-\sum_r L(r) \sum_X u_X(r) \varphi_X(r) + F^{\rm int}(\{\varphi\}) \\
- \sum_r L(r) \alpha(r) \left ( 1-\sum_X \varphi_X(r)\right) \nonumber 
\label{GibbsSF}
\end{eqnarray}
Here and below we normalize all energies by the thermal energy $kT$. The first term on the rhs of this equation features the mean-field partition function $Q$, which can be computed once the segment potentials are known. In the mean-field approximation it is composed of single-chain partition functions:
\begin{equation}
Q = \Pi_X \frac{(q_X(\{u\}))^{n_X}}{(n_X)!}
\label{Q}
\end{equation}
Where $q_X$ is the single-chain partition function of the molecule type $X$, which in turn can be computed once the segment potentials are available. To compute this quantity it is necessary to specify the chain model. Below we will go in more detail. In Eq. \ref{Q} the variable $n_X$ is the number of molecules of type $X$ in the system.

The third term on the rhs of Eq. \ref{GibbsSF} expresses the free energy of interaction in the system. Again, we will go in more detail below. Here it suffices to mention that it can be evaluated once the volume fractions are known.
 
The second term on the rhs transforms the free energy which is specified in the potential domain (as expressed by the first term) to the classical free energy in the ($n,V,T$) ensemble. 

The fourth term implements the incompressibility constraint for each coordinate. In other words, the value of the Lagrange field $\alpha(r)$ is coupled to the requirement that on each coordinate the volume fractions add up to unity. In passing we note that in the incompressible system there is no volume work and the Gibbs energy is the same as the Helmholtz energy. 

Eq. \ref{GibbsSF} has dependences on the segment volume fractions, the segment potentials and the Lagrange field. The free energy as expressed in Eq. \ref{GibbsSF} not automatically has physical significance: it needs to be minimized with respect to the volume fractions and maximized with respect to the segment potentials and the Lagrange field. In equations we are looking for the so-called SF-SCF point for which: 

\begin{equation}
\frac{\delta G}{\delta \varphi_X(r)} = 0 
\label{SCFa} 
\end{equation}
\begin{equation}
\frac{\delta G}{\delta \alpha(r)} = 0
\label{SCFc} 
\end{equation}
\begin{equation}
\frac{\delta G}{\delta u_X(r)} = 0
\label{SCFb} 
\end{equation}

Eq. \ref{SCFa} leads to an expression for the segment potentials in term of the volume fractions. Here we take a Flory-Huggins \cite{BookFleer} type Ansatz, wherein only nearest-neighbor interactions are accounted for. It implements the Bragg-Williams approximation \cite{Hill1962} and use Flory-Huggins interaction parameters $\chi$ to specify the strength of the interactions which has non-trivial values for the unlike contacts only
\begin{equation}
u_X(r) = \alpha(r) + \sum_Y {\chi_{XY}} \left ( \left \langle \varphi_Y(r) \right \rangle - \varphi_Y^b   \right )
\label{SCF_u}
\end{equation}
where the summation runs again over all segment types and $\varphi_Y^b$ is the volume fraction of segments of type $Y$ in the bulk.  

Eq. \ref{SCFc} enforces that system obeys the compressibility constraint, that is:
\begin{equation}
\sum_X \varphi_X(r) = 1
\end{equation}

Last, but not least Eq. \ref{SCFb} leads to the rule to compute the volume fractions from the potentials. Formally the result is 
\begin{equation}
\varphi_X(r) = -\frac{1}{L(r)}\frac{\partial \ln Q}{\partial u_X(r)}
\label{phi}
\end{equation}
The computation of the functional derivative $\partial \ln Q / \partial u_X(r)$ is, in general, rather hard. For a freely-jointed chain, however, there exist an extraordinary efficient propagator formalism which exactly computes the volume fraction as specified by Eq. \ref{phi}. This formalism is outlined in the next paragraph. 

\subsubsection{The propagator formalism}
\label{sec:4.2.4}
Motivated by the close analogy between the diffusion of a Brownian particle and the flight of a random walk, there exist a diffusion-like equation to evaluate the partition function of Gaussian chains. Edwards \cite{Edwards1965} realized that the difference between the diffusion process and the polymer chain is that the polymer cannot visit previously occupied sites. This is known as the excluded-volume problem. He came up with a modified diffusion equation, which corrects, in first order, for the volume interactions which in spherical coordinates reads:
\begin{equation}
\frac{\partial G}{\partial s}= \frac{1}{6}\left (\frac{1}{r^2} \frac{\partial }{\partial r} r^2 \frac{\partial G}{\partial r} \right ) - u G
\end{equation} 
which must be supplemented with initial and boundary conditions. The quantity $G = G(r;s)$  which obeys to this differential equation is related to the partition function and can be used to compute the volume fraction distribution for a given chain molecule. We map this differential equation onto the lattice. Here we cannot go in full details and discuss the resulting formalism instead. By implementing it, the chain model changes from the Gaussian chain to the freely jointed one. The fundamental difference being that formally the chain ends can be separated beyond the contour length in the Gaussian chain, whereas it is not possible in the freely jointed model (finite extensibility).

Let's introduce the free segment distribution function for a segment type $X$ as $G_X(r)= \exp [-u_X(r)]$, which is the Boltzmann weight for a segment $X$ at location $r$. We generalize this quantity to the chain type $i$ and ranking number $s$ dependent quantity by using the chain architecture operators:
\begin{equation}
G_i(r,s) = \sum_X G_X(r) \delta_{i,s}^X
\end{equation}
We may start for molecule $i$ the propagators by setting the statistical weight for the first segment to the free segment distribution function: $G_i(r;1)=G_i(r,1)$. End point distribution functions for segments $s$ along the chain now depend on similar quantities for segment $s-1$ according to the propagator:
\begin{equation}
G_i(r;s) = G_i(r,s) \left \langle G_i(r;s-1) \right \rangle
\end{equation}
where the angular brackets are defined in Eq. \ref{site}. The end-point distribution of the terminal segment is related to the single-chain partition function: 
\begin{equation}
q_i = \sum_r L(r) G_i(r;N)
\end{equation}
In the general case one has to compute also the complementary end-point distribution functions before the volume fractions can be evaluated. As in our case the triblock copolymers are symmetric we can make use of a shortcut and compute the volume fractions by:
\begin{equation}
\varphi_i(r,s) = \frac{n_i}{q_i}\frac{G_i(r;s)G_i(r;N-s+1)}{G_i(r,s)}
\label{phi(r,s)}
\end{equation}
where a chain fragment with $s$ segments is combined with one with $N - s + 1$ segments. The normalization with the free segment distribution is needed to correct for the fact that both walks already have the statistical weight for the connecting segment.

In passing we note that the normalization in Eq. \ref{phi(r,s)} can be used to evaluate the volume fractions in the bulk. It can be shown that  $\frac{n_i}{q_i}= \varphi_i^b/N_i$. The volume fractions in the bulk for the various segment types $X$ follow trivially from the bulk volume fractions per molecule. 

The volume fraction profile of the solvent reads:
\begin{equation}
\varphi_S(r) = \varphi_S^b G_S(r)
\end{equation}
wherein the volume fraction of solvent in the bulk is given by $\varphi_S^b = 1-\sum'_X \varphi_X^b$; the prime on the summation sign indicates that in the sum $X=S$ is not included. The latter equation enforces that the bulk is incompressible. 

\subsubsection{The SF-SCF solution and the grand potential}
\label{sec:4.2.5}
The previous paragraph outlined how the volume fractions can be computed from the potentials and Eq. \ref{SCF_u} implemented the evaluation of the potentials from the volume fractions. Numerically we search for the so-called self-consistent field solution for which the potentials and the volume fractions are mutually consistent (we implemented a precision of at least 7 significant digits), while at the same time the incompressibility constraint is obeyed. When this solution is found, which typically takes only a few seconds CPU time on a modern PC, we can evaluate the free energy of Eq. \ref{GibbsSF}. Starting from the free energy we can extract various other thermodynamic potentials. Relevant for self-assembly we should compute the grand potential. It is possible to evaluate the grand potential $\Omega$ from the summation over the grand potential density:  $\Omega=\sum_r L(r) \omega(r)$  and in turn the grand potential density $\omega(r)$ is given by 
\begin{equation}
\omega(r) = -\sum_i \frac{\varphi_i(r)-\varphi_i^b}{N_i} - \alpha(r) - \frac{1}{2}\sum_{X,Y}\chi_{XY}\left ( \varphi_X(r) \left \langle \varphi_Y(r)\right \rangle - \varphi_X^b\varphi_Y^b  \right )
\end{equation}
which may be interpreted as a local tangential pressure in the micelle. 

In the section on the thermodynamics of micelle formation we already mention that the sub-division potential is related to the grand potential $\Omega$. The formal difference between these two quantities is that in the grand potential as found in the SF-SCF model, the translational entropy of the micelle as a whole is not accounted for, whereas in the sub-division potential the translational entropy is part of it. We will consider polymer micelles for which the degrees of freedom in the translation of the micelle as a whole can be ignored with respect to other contributions, and therefore it is reasonable to equate the grand potential to the sub-division potential. Hence, our interest will be in micelles that have a vanishing grand potential only. As this point is rather subtle, we will pay more close attention to the thermodynamic stability of micelles at the start of the results section.

\subsection{The Kuhn lengths}
\label{sec:4.3}
\begin{figure}
	\centering
		\includegraphics[width=12cm]{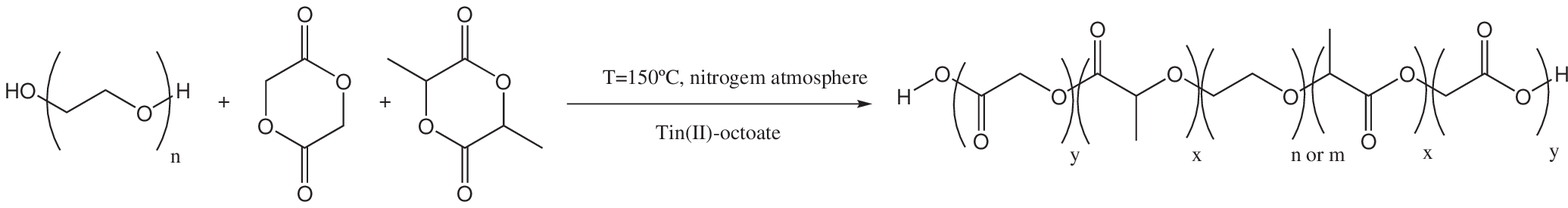}
		\includegraphics[width=12cm]{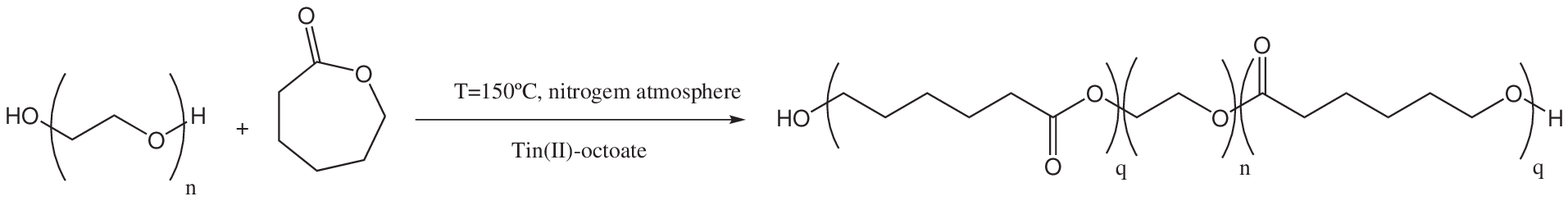}
	\caption{Schematic of the ring-opening polymerization of the triblock copolymers. In our case $x=y$, and $x+y$ is the number of D,L-Lactide and Glycolide repeating units randomly distributed in the hydrophobic end blocks. In the case of $m=136$ ethylene oxide repeating units $x+y$ is $115$, referred to as TBB1, and for $n=68$ ethylene oxide repeating units $x+y$ is $58$, referred to as TBB2 for the PLGA based triblock copolymers. For the PCL based triblock copolymer, referred to as TBC1, there are $n=68$ ethylene oxide repeating units with $q=17$ caprolactone repeating units.}
	\label{fig:20}
\end{figure}

The calculations are targeted to model copolymers with PEO, PCL and PLGA blocks, see Fig. \ref{fig:20}.
For polymeric compounds it is required to describe the chains as Kuhn chains, so that each segment can assume any position in space with respect to the previous segment except for back folding. Since the PEO parts of the copolymers stick into the aqueous solution, and the flexibility of the PLGA and PCL parts do not have a very different chain stiffness chains, we use the Kuhn length of PEO chains, being $b=0.8$ nm \cite{Vincent1990}, as lattice unit in the SF-SCF computations. Each unit in terms of r thus equals 0.8 nm. For instance for a PEO chain with a molar mass of 6k consists of $N=M/M_{mon}=6000/44\approx136$ segments, $M_{mon}$ being 44 g/mol for PEO. Each ethylene oxide monomer has a length of 0.36 nm. Hence a Kuhn segment consists of $0.8/0.32\approx 2.22$ real segments. This means the number of Kuhn segments equals $136/2.22\approx60$ segments \cite{Vincent1990}. In a similar fashion we can estimate the effective number of Kuhn segments for PCL and PLGA blocks and we come to the numbers listed in Table \ref{tab:8}.

\begin{table}[ht]
	\centering 
		\begin{tabular} {|p{5cm}|p{2cm}|} 
\hline
Blocks in copolymers	& $N_K$\\
\hline
PEO 6k	&60\\
PEO 3k	&30\\
PLGA 7.5k	&60\\
PLGA 3.75k&	30\\
PCL 1.9k	&17\\
\hline		
\end{tabular}
	\caption{Number of Kuhn segments ($N_K$) of blocks used in the copolymers studied.}
	\label{tab:8}
\end{table}

\subsection{The Flory-Huggins parameters}
\label{sec:4.4}

 A key moment is to estimate the Flory-Huggins parameters between all components in the mixtures. It must be noted that the values of the interaction parameters should represent the average solvent quality upon solvent exchange upto the point that the micelles become kinetically frozen. It is known that PEO monomers are well-soluble in water at room temperature and often a  $\chi$-parameter of 0.4 is used for describing PEO chains in water \cite{41}. PCL and PLGA are not soluble in water. The $\chi$-parameters of their monomers must be bigger than 0.5. It is also known that the monomers in PCL are more hydrophobic than those in PLGA. Some preliminary calculations resulted in a set of $\chi$-parameter summarized in Table \ref{tab:9_}. We note that we did not try to fine-tune the $\chi$-values and mention that the reasonable comparison with experiments justifies the values used.

\begin{table}[ht]
	\centering 
		\begin{tabular} {|p{5cm}|p{2cm}|} 
\hline
Monomer-solvent interaction		& $\chi$\\
\hline
EG - water	&0.4\\
LGA - water&	1.6\\
CL - water&	3.0\\
LGA - EG&	1.0\\
CL - EG &	1.0\\
\hline		
\end{tabular}
	\caption{$\chi$-Parameters for the monomer-solvent interaction used in SF-SCF computations (EG: ethylene glycol, LGA: Lactic-co-glycolic acid and CL: caprolactone. The block lengths and the corresponding Kuhn lengths are collected in Table \ref{tab:8}.}
	\label{tab:9_}
\end{table}
 
\section{Results}
\label{sec:5} 
\subsection*{Grand potential and equilibrium micelle}
\label{sec:5.1}
The SCF model focuses on the most likely micelle for a system with specified copolymer chain, interaction parameters and concentration in a selective solvent. In the calculations there exists one single micelle in the center of the coordinate system bounded by $M$ spherical lattice layers. For a given number of copolymers in the system, it is possible to compute the aggregation number $g$, defined by the excess number of copolymers in the micelle, i.e., $g = \frac{1}{N_i} \sum_r L(r) (\varphi_i(r)-\varphi_i^b)$. Above it was argued that thermodynamically stable micelles obey to $\varepsilon = 0$. In the SCF model we compute the grand potential $\Omega (g)$ of a micelle that is at the center of the coordinate system and thus the micelle without translational degrees of freedom. For not too concentrated micellar solutions we may write
\begin{equation}
\varepsilon = \ln \varphi_m + \Omega
\label{ss}
\end{equation}
The quantity $\Omega$ is the grand potential of the micelle of which the translational degrees of freedom are frozen as indicated above. Adding $-TS/k_BT=-S/k_BT \approx \ln \varphi_m$ takes into account the mixing entropy. This yields the standard state subdivision potential $\varepsilon$ \cite{Hill1991} which equals zero under equilibrium conditions. Note again that all terms are normalized by $kT$.  Using Eq. \ref{ss} we may compute the volume fraction of micelles from the grand potential, i.e. $\varphi_m (g) = \exp [-\Omega(g)]$. Clearly, $\Omega \ge 0$ or else the micelle volume fraction exceeds unity and clearly micelles with $\Omega >> 1$ can only exist at extremely low micelle concentrations.

\begin{figure}
	\centering
		\includegraphics[width=8cm]{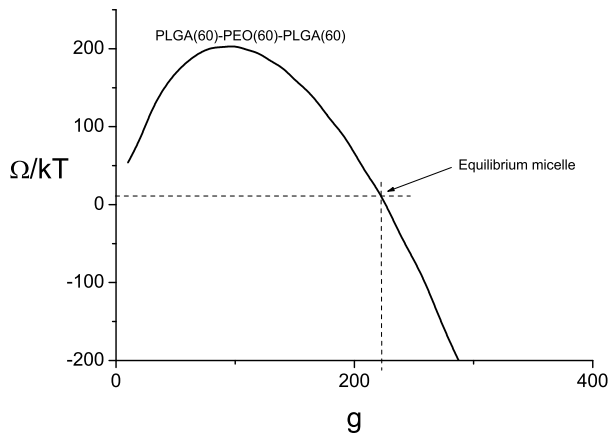}
	\caption{Grand potential of formation of a micelle consisting of PLGA$_{60}$PEO$_{60}$PLGA$_{60}$ of triblock copolymers as a function of the number of copolymers per micelle ($g$).}
	\label{fig:3}
\end{figure}

From the above it is evident that it is necessary to analyze the grand potential $\Omega$  of the micelle as a function of $g$ \cite{Hurter1993,Lyklema2005}. In Fig.\ref{fig:3} we present, as an example, SF-SCF results for the grand potential for a spherical micelle composed with $g$ PLGA$_{60}$PEO$_{60}$PLGA$_{60}$ copolymers. These copolymers contain three blocks of 60 segments each and is described using the  $\chi$-parameters of Table \ref{tab:9_}. For a micelle consisting of just a few copolymers $\Omega$ increases with $g$, analogously to the nucleation of small droplets in an oversaturated solution. These micelles are thermodynamically unstable due to the large surface-to-volume ratio. Indeed as long as $\partial \Omega/\partial g > 0$, the micelle is unstable (free energy has a local maximum), implying that micelles with this aggregation number will have a vanishing low probability. From Fig.\ref{fig:3} it is seen that for $g>100$ the grand potential becomes a decreasing function of $g$. This is the signature of micelles that become thermodynamically stable (free energy has a local minimum).

The first micelles, that is when $\partial \Omega /\partial g=0$, that are stable have an aggregation number $g=g^* \approx 100$ and the concentration in solution for this system may be identified as the CMC. For micelles with $g>g^*$ the grand potential decreases with aggregation number, that is $\partial \Omega/\partial g < 0$. The chains in the corona of the micelles are packed closer and closer to each other and this generates a pressure in the corona that opposes the growth of the micelles. 

In the example of Fig.\ref{fig:3} the micelle concentration at the CMC is exceedingly low. From a practical point of view we should therefore focus on micelles that have a higher aggregation number. In surfactant problems it has been advised to focus on micelles with a reasonable amount of translational entropy, e.g., $\Omega(g) \approx 10$kT, implying a volume fraction of micelles that are still dilute, but measurable by light scattering. 
 	
Using this Ansatz, we extract from Fig. \ref{fig:3} that most-likely micelles consisting of PLGA$_{60}$PEO$_{60}$PLGA$_{60}$ copolymers are composed of on average 237 copolymers. Of course one should expect that in practice there are fluctuations in micelle composition. In other words that micelles with a smaller or larger aggregation number must be expected.
Within the SCF model it is also possible to estimate the width of the micelle size distribution.

\begin{figure}
	\centering
		\includegraphics[width=8cm]{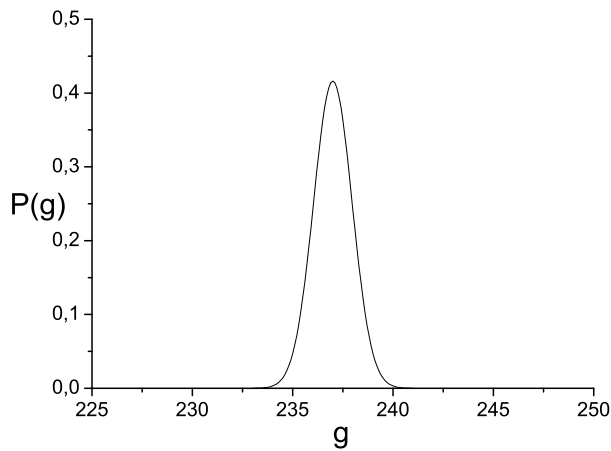}
	\caption{Probability distribution of the number of copolymers per micelle for a micelle of PLGA$_{60}$PEO$_{60}$PLGA$_{60}$ triblock copolymers.}
	\label{fig:4}
\end{figure}

From statistical thermodynamics it follows that $\partial g/\partial \mu = \left< g^2\right>-\left<g \right>^2 = \delta g$, often referred to as the dispersion of fluctuations (in our case fluctuations in the micelle size), wherein $\mu$ is dimensionless. It can be shown that the SCF equations obey the Gibbs-Duhem relation 
$$\partial \Omega / \partial \mu = -g \quad.$$
Multiplication of both sides with ${\partial \mu}/{\partial g}$ results in
\begin{equation}
\frac{\partial \Omega }{\partial g} = - \frac{g}{\delta g} \quad.
\end{equation}
We give the resulting micelle size distribution in Fig.\ref{fig:4}, assuming a Gaussian size distribution. The polydispersity, as predicted by the SCF model, is very narrow; the standard deviation is just 4\%. It should be realized that SF-SCF is based on a mean-field approach in which shape fluctuations are not accounted for and therefore we expect that the size distribution is somewhat underestimated. As compared to the experimental counterpart we further expect that the theory underestimates the width of the size distribution because in the experimental samples the polymers are both polydisperse in the overall molecular weight as well as with respect to the block sizes. A more detailed SCF analysis can be implemented to account for a distribution of chain lengths. Here we can not do this because the detailed information about the distributions is not yet available. 
 	
\begin{figure}
	\centering
		\includegraphics[width=8cm]{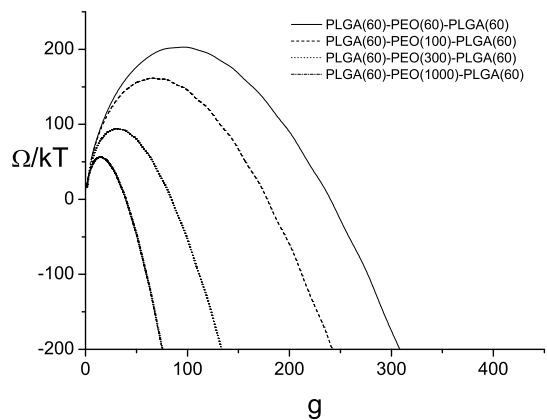}
	\caption{Grand potential of formation of a micelle consisting of PLGA$_{60}$PEO$_{N_X}$PLGA$_{60}$  triblock copolymers with varying PEO chain lengths $N_X$ for $N_X=60$ and larger.}
	\label{fig:5}
\end{figure}

In Fig.\ref{fig:5} we demonstrate what happens when the PEO block length is increased, while keeping the chain lengths of the PLGA blocks fixed. It follows that the equilibrium number of copolymers per micelle is decreasing with increasing chain length of the PEO block. This effect can easily be rationalized considering the packing arguments. The outside of the copolymer micelle must be covered with solvophilic polymer blocks being PEO. Obviously, a certain amount of PEO on the outside of the micelles is required in order to provide sufficient stability. As the PEO block length increases at given $g$ there is more PEO on the outside of the micelle. Hence $g$ can be lowered to maintain the same stability of a micelle.
 	 
\begin{figure}
	\centering
		\includegraphics[width=8cm]{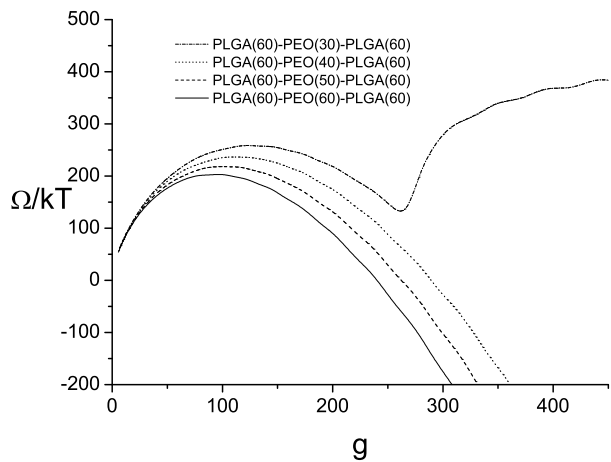}
	\caption{Grand potential of formation of a micelle consisting of PLGA$_{60}$PEO$_{N_X}$PLGA$_{60}$  triblock copolymers with varying PEO chain lengths $N_X$ for $N_X=60$ and smaller.}
	\label{fig:6}
\end{figure}

In Fig. \ref{fig:6} we show stability curves for the case of decreasing the PEO chain length. The most like micelle size, that is $g$-value, increases with decreasing PEO chain length, as can be expected from the results in Fig. \ref{fig:5}. When the length of the PEO moiety is decreased there exists a limit below which spherical micelles can no longer find their tensionless state. This is illustrated in Fig. \ref{fig:6}. When the length of PEO is decreased towards a value of 30 segments, the grand potential does not drop to values near $\Omega=0$, but start to increase with $g$ above some $g> g^{**}$. This implies that the theory predicts that there is an upper limit in the micelle concentration. Spherical micelles with $g>g^{**}$ are unstable and micelles with a cylindrical or lamellar topology are expected instead. In other words, the overall composition of the copolymers is simply too solvophobic to self-assemble in stable spherical micelles.

Obviously, there is a limit to the composition of the block copolymers that can self-assemble into spherical micelles. We have rationalized this limit in a patent application draft \cite{patent2011}.

\subsection{Radial density profiles and micelle size}
\label{sec:5.2}
\begin{figure}
	\centering
		\includegraphics[width=8cm]{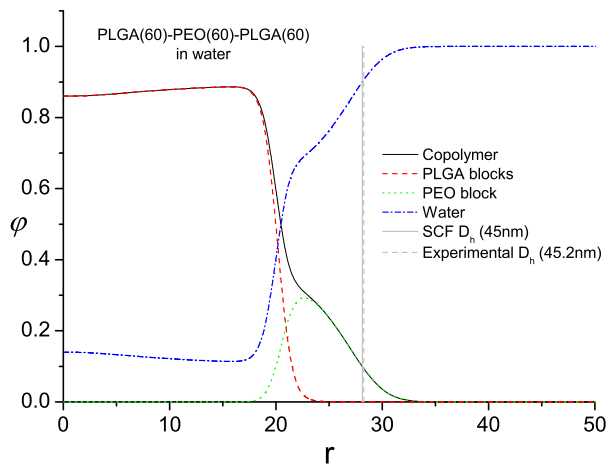}
	\caption{Equilibrium radial density profiles of water, total copolymer, PLGA blocks and PEO block as a function of the center from a micelle $r$. The micelle consists of PLGA$_{60}$PEO$_{60}$PLGA$_{60}$  triblock copolymers.}
	\label{fig:7}
\end{figure}

Once the most-likely number of copolymers $g$ in the micelle is known, the radial segment density profiles of all components composing the spherical micelle can be analyzed. In Fig. \ref{fig:7} we show the radial density profiles of copolymer segments, solvent, and separate PEO and PLGA blocks for a PLGA$_{60}$PEO$_{60}$PLGA$_{60}$ triblock copolymer micelle that corresponds to using 7.5k - 6k - 7.5k PLGA-PEO-PLGA triblocks, see Table \ref{tab:9}. This micelle consists of 237 copolymers, see Table \ref{tab:10}. The density profile as a function of the radial distance $r$, commences at $r=0$, the center of the core towards large $r$ values, far from the micelle. It is noted that $r$ is given in lattice units. Each lattice unit thus corresponds to 0.8 nm; the Kuhn length for PEO.

\begin{table}[ht]
	\centering 
		\begin{tabular} {|p{2cm}|p{2cm}|p{2cm}|p{2cm}|} 
\hline
Copolymer ID	& $D_{hp}$ (nm) & $D_{ht}$ (SF-SCF) (nm) & $|\Delta I|$ \\
\hline
TBB1	&45.2	&45	&0.44\\
TBB2	&31.3	&28	&10.54\\
TBC1	&27.2	&26	&4.41\\
\hline		
\end{tabular}
	\caption{Comparison of experimental and theoretical SF-SCF hydrodynamic diameters $D_h$ of copolymeric micelles prepared using the nanoprecipitation method using copolymers of compositions as indicated.
	$D_{ht}$ = Theoretical hydrodynamic diameter, according to SF-SCF. $D_{hp}$ = Practical hydrodynamic diameter measured by DLS \cite{LebouilleA2011}. $|\Delta I|$ is the percentual deviation between what is practically measured and theoretically calculated. For copolymer ID see Table \ref{tab:10}.}
	\label{tab:9}
\end{table}

In the center of the micelle, or core, there is a nearly constant volume fraction of copolymers of (in this case) about $\varphi \approx 0.87$  and about 13 vol\% of water molecules, also confirmed in various other publications \cite{Goldmints1999,King1997,Linse1993,Pedersen2003}. The amount of water is substantial for mixing entropy reasons: full exclusion of water is unlikely as this costs a lot of mixing entropy. The $\chi$-value between PLGA monomers and water molecules is taken as 1.6, which causes demixing, but is not that extreme. Indeed the core will dry up with increasing $\chi$-value. The slight increase of water towards the core is caused by the presence of more PLGA end segments in the core. Near such end groups it is somewhat less unfavourable to have water molecules. Moreover the chains have to stretch to reach the micelle center. By having slightly more solvent in the core, the stretching of the chains can be reduced somewhat.  

Around $r=20$ the water concentration (dash dot line) increases significantly and the copolymer concentration drops correspondingly. The distribution of the PLGA (dash line) and PEO blocks are also plotted (dot line). The hydrophobic PLGA monomers are in the core, while the PEO segments are completely expelled from the core and are all located in the micellar corona. The PEO density goes through a maximum of about 25 vol\% of segments providing steric stabilization. A rough estimation of the size can already be made based on these density profiles. Near $r=30$ the density profile of copolymer segments drops to such low values that these can not be seen in these coordinates. This means an effective radius of about 30 times 0.8 nm= 24 nm or a diameter of 48 nm.

\begin{table}[ht]
	\centering 
		\begin{tabular} {|p{3cm}|p{3cm}|p{3cm}|p{0.6cm}|p{2.2cm}|} 
\hline
Copolymer	& Molar masses (1k = 1000 g/mol) & Polymer composition& $g$&  Copolymer ID \\
\hline
PLGA-PEO-PLGA&	3.75k - 3k - 3.75k&	30-30-30&	132& TBB2\\
PLGA-PEO-PLGA&	7.5k - 6k - 7.5k&	60-60-60&	237& TBB1\\
PCL-PEO-PCL&	1.9k - 3k - 1.9k&	17-30-17&	155& TBC1\\
	\hline		
\end{tabular}
	\caption{SF-SCF determined averaged number of copolymers per micelle $g$.}
	\label{tab:10}
\end{table}

Since we measure the averaged hydrodynamic diameter $dH$ using dynamic light scattering we also computed the hydrodynamic diameter of the micelles using Brinkman-Debye theory \cite{CohenStuart1984,Scheutjens1986}, for which the copolymer density profile is needed as input. The resulting values for $D_h$ are plotted in Fig. 3 of the companion publication \cite{LebouilleA2011}, so $D_h=45$ nm for the copolymer density profile in Fig. \ref{fig:7}. This value corresponds very well to the predicted hydrodynamic diameter given in Table \ref{tab:9}. In view of polydispersity effects and uncertainties in estimating the properties of the micelles for the SF-SCF computations we may only claim that the micellar size can be well predicted. It seems therefore that the micelles can be described by an theory that focuses on equilibrium structures. 
 
\begin{figure}
	\centering
		\includegraphics[width=8cm]{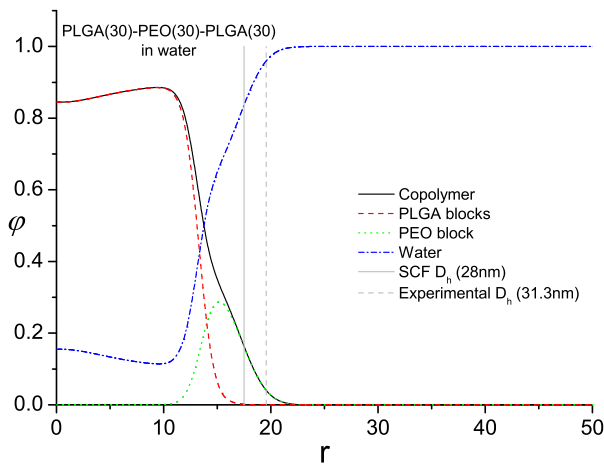}
	\caption{Equilibrium radial density profiles of water, total copolymer, PLGA blocks and PEO block as a function of the center from a micelle $r$. The micelle consists of PLGA$_{30}$PEO$_{30}$PLGA$_{30}$ triblock copolymers.}
	\label{fig:8}
\end{figure}

In Fig. \ref{fig:8} we show the radial density profiles of a PLGA$_{30}$PEO$_{30}$PLGA$_{30}$ triblock copolymer micelle composed of 3.75k - 3k - 3.75k PLGA-PEO-PLGA triblocks (Table \ref{tab:9}). This micelle consists of 132 copolymers (Table \ref{tab:10}). 

In comparison to the profile in Fig. \ref{fig:7} for the micelle with larger copolymers we observe the density profile inside the core of the micelle varies more strongly. Also the size of the micelles is smaller as can be expected; the hydrophobic chains are only 30 segments long, so the cores are smaller, and the stabilizing PEO chains on the outside are smaller as well. As a rough estimation one might speculate that $g$ is half of the value for micelles composed of PLGA$_{60}$PEO$_{60}$PLGA$_{60}$ micelles ($g=237$). Indeed the $g$ value of 132 is only a bit larger than an estimated 119. As a consequence the size should in a naive picture scale as $d_1 \approx d_2(1/2)^{1/3}$, implying a diameter of about 36 nm for $d_2=45$ nm. Still, the SF-SCF size of 28 nm (Table \ref{tab:9}) is even smaller. Hence more copolymer as expected is needed to stabilize a smaller particle as the molar mass decreases.

\begin{figure}
	\centering
		\includegraphics[width=8cm]{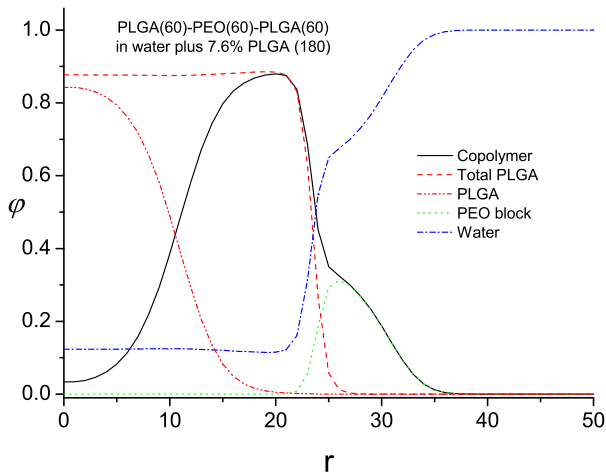}
	\caption{Equilibrium radial density profiles of water, total copolymer, PLGA blocks and PEO block as a function of the center from a micelle $r$. The micelle consists of  PLGA$_{60}$PEO$_{60}$PLGA$_{60}$ triblock copolymers with added free PLGA$_{180}$ copolymers.}
	\label{fig:9}
\end{figure}

Next, we discuss the effect of encapsulation of hydrophobic compounds in the triblock copolymer micelles. Here we choose PLGA (homopolymer) chains with a molar mass of 20k , corresponding to 180 segments, as the guest molecules. These will be fairly insoluble in the aqueous bulk and will prefer to be encapsulated in the core of the micelle because of the PLGA environment. The composition of the micelle with 7.6\% encapsulated free PLGA in a micelle composed of identical copolymers as in Fig. \ref{fig:7} is plotted in Fig. \ref{fig:9}. The number of copolymers per micelle now increased from 237 to 337 and the diameter increases from 45 to 51 nm. As expected we see that, whereas the PLGA monomer concentration in the core is constant, the free PLGA is more concentrated close to the centre of the micelle and the PLGA monomers connected to the triblocks concentrate in the outer core region. Not shown is the finding that the influence of molar mass of the free PLGA is nearly imperceptible. SF-SCF obviously enables to study encapsulation effects and efficiencies. Once all Flory-Huggins $\chi$-parameters are known between any drug molecule, the polymer segments and the solvent, SF-SCF allows studying encapsulation equilibriums. This computation inspired us to make the study leading to the results that will be presented in the companion publication \cite{LebouilleA2011}.

\begin{figure}
	\centering
		\includegraphics[width=8cm]{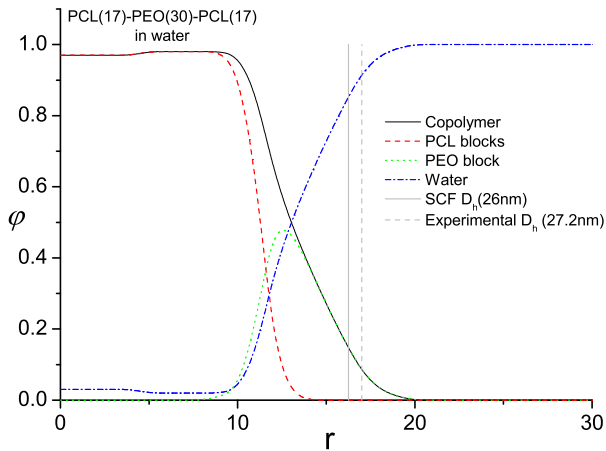}
	\caption{Equilibrium radial density profiles of water, total copolymer, PEO blocks and PCL block as a function of the center from a micelle $r$. The micelle consists of PCL$_{17}$PEO$_{30}$PCL$_{17}$ triblock copolymers.}
	\label{fig:10}
\end{figure}

We have also studied triblocks with the hydrophobic polymer PCL, replacing PLGA. In Fig. \ref{fig:10} we have plotted SF-SCF results for a micelle composed of PCL$_{17}$PEO$_{30}$PCL$_{17}$ triblock copolymers using 1.9k - 3k - 1.9k PCL-PEO-PCL triblocks (see again Table \ref{tab:9} for the  $\chi$-parameters used). For this micelle we find it consists of 155 copolymers per micelle (Table \ref{tab:10}). Since the  $\chi$-parameter is estimated to be substantially larger (3.0) the core now hardly contains water and can merely be viewed upon as a PCL melt environment. In the corona the PEO again goes through a maximum volume fraction that now reaches a maximum value of nearly 50 vol\% of PEO segments. It seems the PEO segments here screen the hydrophobic core more strongly.  They interact as a 'mediator' between water and the very hydrophobic core and in this case the peak is more sharp due to a more hydrophobic core environment. This might also have consequences for drug release; once the drugs leave the hydrophobic core the drugs need to pass the PEO barrier before they are released from the micelle.

In summary, we have shown that the SF-SCF theory may be used as a tool to unravel the structure-function relationship between copolymer composition and micellar size and morphology, also for situations that the resulting micelles are structurally quenched. Hence, using SF-SCF predictions allow for more efficient experimentation. As discussed more thoroughly in the companion publication \cite{LebouilleA2011}, by utilizing this approach we were able to prepare nanosized particles consisting of PLGA-PEO-PLGA (7.5-6-7.5 and 3.75-3-3.75) or PCL-PEO-PCL (1.9-3-1.9) block copolymers in which (several) hydrophobic compounds can be encapsulated. One of the reasons to do this is that Ostwald ripening was minimized. Stabilization of micelles by block copolymers prevents particle aggregation, but the stabilizing polymer layer is open enough to allow solute mass transfer. In order to prevent/minimize solute transfer it is desired to tune the particle core composition to prevent this mass transfer. Additionally, the solubility of the encapsulated compound can be decreased by antisolvent addition to the bulk resulting in a significant slow down of Ostwald ripening. The extremely low solubility of the used triblock copolymers limits copolymer exchange between micelle and bulk again minimizing solute mass transfer and slowing down Ostwald ripening. There is no need to use surfactant in this process, conventional nanoprecipitation processes need an excess of surfactant, mostly very water soluble with relative high CMC's. Since we incorporated the surfactant function in the polymer backbone no exchange of adsorbed and free surfactant is needed for stable suspensions. This also avoids washing the nanoparticle suspension to remove excess of free surfactant used in the process and limits Ostwald ripening. We were able to synthesize different kinds of triblock copolymers allowing simultaneous tuning of the size and loading. When performing the nanoprecipitation process there is hardly an influence of temperature and triblock copolymer molar mass polydispersity. However, using these micelles in electrolytes, e.g. in vivo, care must be taken to avoid destabilization of the micelles due to electrostatic interactions. Non reported data shows that it is feasible to perform the nanoprecipitation process, using the mentioned triblock copolymers, in different electrolytes at different pH's and that the suspension stays stable in time.
 
\section{Conclusions}
\label{sec:7}
In summary, we have shown that SF-SCF predictions provide an accurate prediction of structural properties of micelles processed via nanoprecipitation and composed of PCL-PEO-PCL and PLGA-PEO-PLGA copolymers. The hydrodynamic size that follows from these computations matches surprisingly well with the measured particle sizes from dynamic light scattering. From the computations it follows that the size of the nanoparticles is determined by the number-averaged molar mass of the block copolymers; polydispersity hardly affects the size of the micelles. We may speculate about reasons why an equilibrium theory can be used for an intrinsically off-equilibrium micelle formation process. One must realize that in the micelle formation procedure the solvent quality goes from a good solvent to a selective solvent. We may suggest that this solvent exchange is sufficiently slow so that chains can respond for some time to a local equilibrium, which we can mimic using some effective (intermediate) parameters. When the solvent quality subsequently becomes more extreme, the cores solidifies and the aggregation number is quenched. The latter may occur relatively suddenly in the process, so that the chains effectively cannot respond to these more selective solvent conditions. The prediction of the aggregation number corresponding to the quench point is apparently possible using a set of effective interaction parameters. For given aggregation number, the theory can then predict accurate radial distribution functions and corresponding hydrodynamic sizes. Modifications of the nanoprecipitation method, e.g. by changing the initial solvent quality and/or the exchange time for the solvent going from good to selective, is expected to have an influence on the best values for the interaction parameters that should be used in subsequent SCF modeling. However, once calibrated for given process conditions, one can proceed also for these new conditions to predict by the SCF theory a value for the aggregation number, the hydrodynamic size and loading capacities.

\end{document}